\documentstyle[11pt,aaspp4,epsfig]{article}
\newcommand{\etal} {et~al.}

\newcommand{\psr}{PSR\,B1937$+$21}
\def\spose#1{\hbox to 0pt{#1\hss}}
\newcommand\lsim{\mathrel{\spose{\lower 3pt\hbox{$\mathchar"218$}}
     \raise 2.0pt\hbox{$\mathchar"13C$}}}
\newcommand\gsim{\mathrel{\spose{\lower 3pt\hbox{$\mathchar"218$}}
     \raise 2.0pt\hbox{$\mathchar"13E$}}}

\input epsf

\begin{document}

\title{Multifrequency Observations of Giant Radio Pulses from the Millisecond 
Pulsar B1937+21}

\author{A. Kinkhabwala}
\affil{Joseph Henry Laboratories and Department of Physics, Princeton
  University, \\Princeton, NJ 08544; ali@princeton.edu}

\and
\author{S. E. Thorsett\altaffilmark{1}}
\affil{Department of Astronomy and Astrophysics, University of California,
\\Santa Cruz, CA 95060; thorsett@ucolick.org}
\altaffiltext{1}{Alfred P. Sloan Research Fellow}


\begin{abstract}

Giant pulses are short, intense outbursts of radio emission with a
power-law intensity distribution that have been observed from the Crab
Pulsar and \psr.  We have undertaken a systematic study of giant pulses from
\psr\ using the Arecibo telescope at 430, 1420, and 2380~MHz.  At 430~MHz,
interstellar scattering broadens giant pulses to durations of 
$\sim50~\mu$secs, but at higher frequencies the pulses are very short, 
typically lasting only $\sim1$--$2~\mu$secs.  At each frequency, giant pulses 
are emitted only in narrow ($\lsim10~\mu$s) windows of pulse phase
located $\sim 55$--$70~\mu$sec after the main and interpulse
peaks.  Although some pulse-to-pulse jitter in arrival times is
observed, the mean arrival phase appears stable; a timing analysis of
the giant pulses yields precision competitive with the best average
profile timing studies.  We have measured the intensity distribution of the 
giant pulses, confirming a roughly power-law distribution with approximate 
index of $-1.8$, contributing $\gsim0.1\%$ to the total flux at each frequency.
We also find that the intensity of giant pulses falls off with a slightly 
steeper power of frequency than the ordinary radio emission.

\end{abstract}


\keywords{pulsars--pulsars: individual (\psr, B0531+21)}
 

\section{Introduction}

Despite their remarkable frequency stability, most radio pulsars exhibit
considerable pulse-to-pulse intensity fluctuations.  A histogram of 
the pulse intensity typically shows a roughly exponential or Lorentzian
shape (e.g., \cite{mt77}), with a tail extending out to perhaps ten times
the mean pulse strength.

The Crab pulsar (PSR~B0531+21), by contrast, exhibits frequent, very strong 
radio pulses extending to hundreds of times the mean pulse intensity (e.g., 
\cite{lcu+95}).  These so-called ``giant pulses'' exhibit a characteristic 
power-law intensity distribution---a phenomenon that for many years was 
observed in no other pulsar.  Despite hints from early observations 
(\cite{wcs84}), it came as a surprise to find qualitatively similar behavior 
from \psr, the first (and still fastest known) millisecond pulsar 
(\cite{sb95,cstt96}).

The Crab pulsar and \psr\ could hardly be less similar in their properties.
The Crab pulsar, born in A.D.~1054, is the youngest known pulsar, and
B1937+21 is one of the oldest, with a characteristic spin-down age of
$\sim2\times10^8$~yrs. The inferred surface magnetic field strength of
the Crab pulsar, $B=4\times10^{12}$~G, is about $10^4$ times as high as
that of B1937+21.  Although the Crab is the fastest of the high field pulsars, 
with a period of $P=33$~ms, it is twenty times slower than the 1.56~ms 
B1937+21.  The pulsars' only identified common feature is their similarly 
strong magnetic field strength at the velocity-of-light cylinder, $\sim10^6$~G,
which is higher than that for any other known pulsar (\cite{cstt96}).
Whether this is coincidence, or in some way responsible for the observed
giant pulse behavior, is unknown.

All previous studies of giant pulses from B1937+21 have been at or near 
430~MHz with the exception of some limited, inconclusive data at 1.4~GHz 
(Wolszczan \etal\ 1984).  We present below the first
multifrequency study of 
giant pulses in B1937+21.  In \S\ref{sec:obs}, we describe the 
observations and giant pulse search algorithms.  A comparison
of the arrival times of normal and giant pulse emission is discussed in 
\S\ref{sec:arr}.  Following this, we discuss pulse morphology in 
\S\ref{sec:mor} and the limits of accurate timing in \S\ref{sec:tim}.  Using 
the arrival data, we present intensity distributions of the giant pulses and 
the contaminant noise in \S\ref{sec:int}.  In \S\ref{sec:spec}, we discuss the 
approximate spectrum of the largest giant pulses.  Finally, in \S\ref{sec:dis},
we point to open questions and necessary future observations.

\section{Observations and Signal Processing}\label{sec:obs}

All observations were made using the Princeton Mark~IV instrument
(\cite{sst+99})  at
the 305~m Arecibo telescope, between 1998 February 21 and 1999 August
1.  Observations were made at three frequencies, as part of an ongoing
timing study of this pulsar.  The pulsar signal strength varied
because of interstellar scintillation; observations made during times
of strong signal were retained for the giant pulse analysis.  The
data used for this work included 30~minutes ($\sim10^6$~pulses) at 430~MHz,
4~hours ($\sim10^7$~pulses) at 1420~MHz, and 26~minutes
($\sim10^6$~pulses) at 2380~MHz.  

After completion of the observations, the data were coherently
dedispersed in software by convolution with a complex chirp function
(\cite{hr75}), to remove the progressive phase delays as a function of
frequency caused by free electrons in the interstellar medium (ISM).
The analysis pipeline has been described by \cite{sst+99}.
At the dispersion measure of \psr (71), with a 10~MHz bandwidth,
the dispersive smearing at 430~MHz, 1420~MHz,
and 2380~MHz totals 74.16~ms, 2.06~ms, and 0.44~ms, respectively, before
coherent dedispersion.  
After dedispersion, the signals from the two orthogonal polarizations were
squared and cross multiplied to produce the four Stokes parameters.
Because we were analyzing data taken primarily for other purposes,  in
most cases insufficient calibration data were available for high
precision polarization calibration.
We therefore concentrate on analysis of the total intensity data.

After coherent dedispersion, the standard timing analysis pipeline was
used to fold the data synchronously with the known
topocentric period of \psr, to produce average profiles.
In a parallel analysis, 
the data were searched for strong individual pulses.  
Initial exploratory analysis confirmed that all strong pulses were 
confined to fairly narrow windows on the tails of the main pulse (MP) and 
interpulse (IP). At 430~MHz, giant pulses were therefore identified
as in Cognard \etal\ (1996), by measuring the integrated flux density in 
150$~\mu$sec windows located on the tails of the 
MP and IP.  At higher frequencies, where the giant pulses were much narrower 
(many lasting only a few $0.1$--$0.2~\mu$sec bins) and appeared in a region of 
pulse phase that was significantly wider than the individual pulses,
giant pulses were identified by 
searching for pairs of bins with combined
energy greater than a threshold level.  Because of the relative computational
efficiency of this procedure, we did not limit our search to particular
regions of pulse phase.  But as at 430~MHz, giant pulses were
only detected in the tails of the MP and IP. (Note that this appears to be
in conflict with the results of Wolszczan \etal\ (1984).)

Interstellar scintillation strongly modulated the apparent intensity of the
pulsar signal from day to day.  Scintillation should affect the
normal and giant pulse emission identically.  Therefore, in order to
compare giant pulse intensities observed on different dates,  we 
accounted for these variations by calibrating the average normal emission flux 
density for each run to the power-law model given by Foster, Fairhead, and 
Backer (1991).  They found that the flux density as a function of
frequency could be expressed as $F$[mJy]$=(25.9\pm2.6)\,\nu^{-2.60\pm0.05}$, 
where $\nu$ is in GHz.  We use units of [Jy] for flux density and 
[Jy$\cdot\mu$sec] for integrated flux density.

\section{Giant pulse distribution in time and pulsar phase}\label{sec:arr}

An important difference between giant pulses in the Crab pulsar and in
\psr\ appears in the distribution of the pulses with respect to the
star's rotational phase.  Individual Crab giant pulses arrive at phases 
distributed throughout most of the emission envelope of the normal pulse
profile.  In contrast, the early observations of giant pulses from B1937+21
at 430~MHz found that they arrived only in narrow regions on the tails of the 
two normal pulse components (\cite{bac95,cstt96}). With our greater 
sensitivity, we confirm this result at 430~MHz and find the same behavior at 
1420 and 2380~MHz.  

\begin{figure}[c]
\centerline{\psfig{file=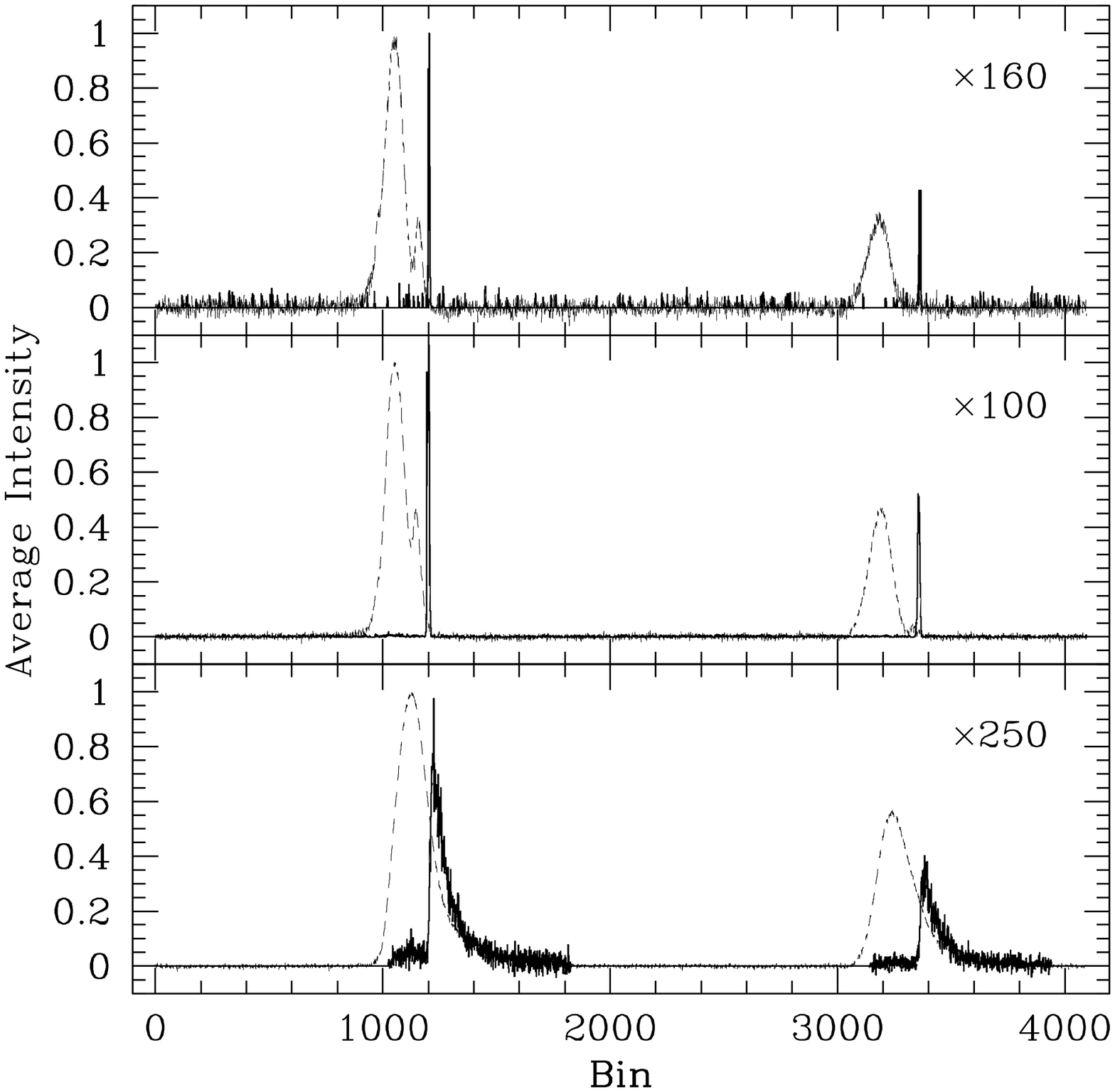,width=3.25in}}
\caption{Folded giant pulse emission (solid) and normal emission (dashed) at 
2380~MHz (top), 1420~MHz (middle), and 430~MHz (bottom) with unit 
intensities of 0.07~Jy, $0.24~$Jy, and $3.0~$Jy (calculated using the 
power-law model in Foster \etal\ (1991)) and 
total integration times 
at each frequency of 26~minutes, 4~hours, and 
30~minutes, respectively.  One pulse period equals 4096 bins with each bin 
$\sim0.38~\mu$secs.  The 2380~MHz profile has been rotated so that the 
MP peak occurs in the same bin as for the 1420~MHz profile.  The 430~MHz 
profiles have been rotated so that the IP peak arrival bin (with scattering 
taken into account) is coincident with that of the 1420~MHz IP peak.  The 
giant pulse profiles have been amplified by the stated amounts.  (Because of 
differences in receiver noise levels and the giant pulse selection threshold, 
a comparison of the amplification factors is not physically meaningful.)
\label{fig:toothpick}}
\end{figure}

Figures \ref{fig:toothpick}--\ref{fig:toothpickIP} show the average flux 
due to all giant pulses as a function of pulse phase along with the average 
normal emission.  For each run, giant pulses were selected using an intensity 
threshold chosen to minimize noise contamination.  As is clear from these 
figures, the giant pulses occur well after the normal emission phase; in 
particular the giant pulses do not produce the ``notch'' emission on the 
trailing edge of the MP as had been speculated by Cognard \etal\ (1996).

To characterize the average properties of the giant pulse emission, we
have fit the average giant pulse profile at each frequency with a Gaussian 
model at 2380 and 1420~MHz and, to account for interstellar scattering, with 
a model consisting of a Gaussian convolved with an exponential tail at 
430~MHz.  We find best-fitting Gaussian widths (FWHM) of 3$~\mu$s 
(4$~\mu$s) for the MP (IP) giant pulse profile at 2380~MHz and 4.3$~\mu$s 
(4.1$~\mu$s) for the MP (IP) giant pulse profile at 1420~MHz.  At 430 MHz, 
we find that the MP and IP giant pulse profiles can be adequately described as 
Gaussians with widths 6.6$~\mu$s and 6.4$~\mu$s, respectively, convolved 
with an exponential scattering tail with $\tau=28~\mu$s.  This scattering 
timescale $\tau$ is similar to that estimated from measurements of the normal 
profile, confirming that the structure observed in the low frequency average 
giant pulse profile is dominated by propagation effects.  (Because our 
exponential scattering model is probably an oversimplification of the true 
effects of scattering on the signal (e.g., \cite{sbh+99}), our estimates of 
the intrinsic width of the mean giant pulse profile at 430~MHz should be 
considered an upper limit.)

At high frequency, the windows in which giant pulse emission occur are much 
narrower than the average pulse emission windows.  Indeed, they are remarkably 
narrow both absolutely and as a fraction of the pulsar period, each 
corresponding to less than one degree of rotational phase.  They are, we 
believe, the sharpest stable features ever detected in a pulsar profile.  It 
is this property that makes the mean giant pulse emission from \psr\ a 
potentially valuable fiducial point for high-precision timing observations, as 
discussed in \S\ref{sec:tim} below.

\begin{figure}[c]
\centerline{\psfig{file=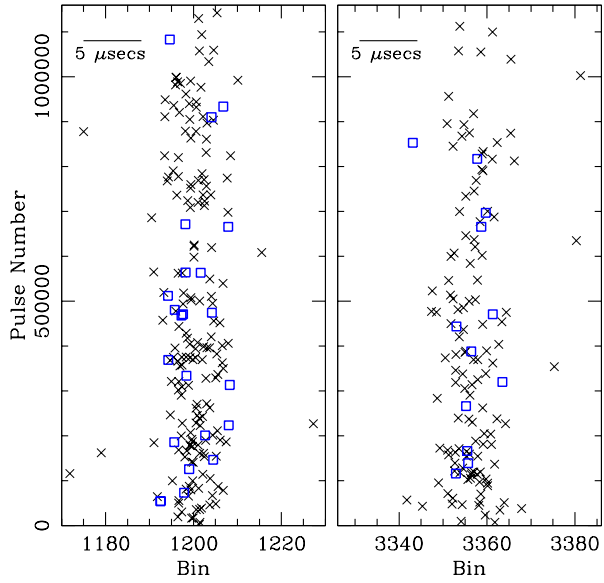,width=3.25in}}
\caption{Distribution of giant pulses at 1420~MHz (on MJD~50865) with
integrated flux density $\geq30~$Jy$\cdot\mu$secs, in both pulse phase (1 bin 
$\simeq0.38~\mu$secs) and pulse number.  Arrival uncertainty for any given 
giant pulse is $\lesssim0.5~\mu$sec.  The left panel shows the MP region, the 
right panel the IP region.  In each region, the pulses are confined to a very
narrow window---10 bins (of 4096 total) corresponds to about $0.9^\circ$ of 
pulse phase---and have a temporal distribution consistent with a Poisson 
process.  The same is true for a subset of the strongest giant pulses, with 
integrated flux density $\geq80~$Jy$\cdot\mu$secs, which are indicated with 
open squares.
\label{fig:arr}}
\end{figure}

We have also investigated the distribution of giant pulses over time, finding 
consistency with Poisson statistics (so neighboring giant pulses appear 
uncorrelated). The distribution in time during one observation is illustrated 
in Fig.~\ref{fig:arr}, which displays all pulses with integrated flux density 
$\geq30~$Jy$\cdot\mu$secs.  Also shown are the brightest pulses, with 
integrated flux density $\geq80~$Jy$\cdot\mu$secs.  The arrival time 
uncertainty for any given giant pulse is $\lesssim0.5~\mu$secs; this will be 
discussed further discussed in \S\ref{sec:mor}.  We find no difference in the 
distributions of the more powerful and less powerful giant pulses, except for 
rate.  The giant pulse distribution also appears very stable over this 
half-hour scan, with no apparent drifting or nulling periods, although the 
density drops off slightly at the end of the scan, especially in the IP, as 
interstellar scintillation reduces the mean pulsar flux relative to our 
threshold values. 

\section{Individual Pulse Morphology}\label{sec:mor}

\begin{figure}[c]
\centerline{\psfig{file=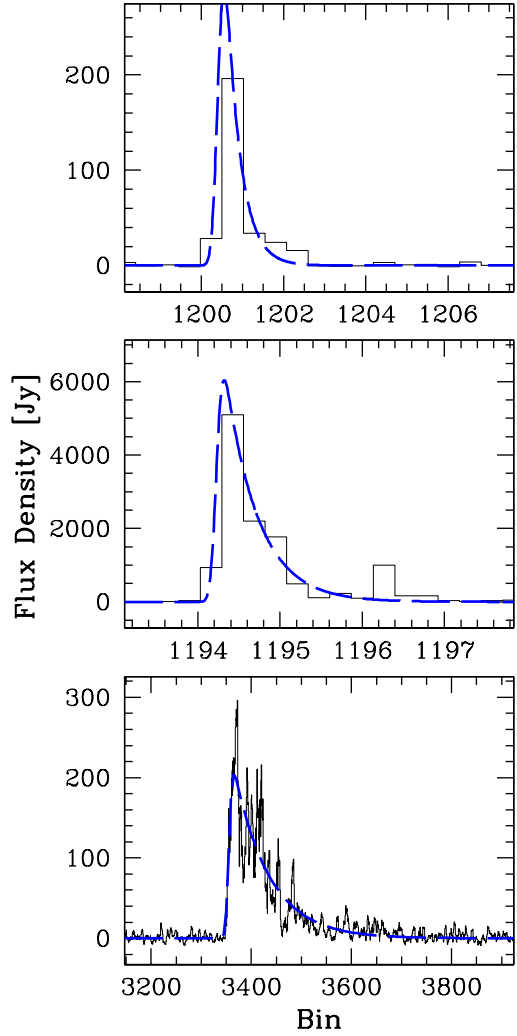,width=6in}}
\caption{Strongest individual giant pulses at 2380 (top), 1420 (middle),
and 430 (bottom)~MHz.  Flux density (solid) and fitted convolved Gaussians 
(long-dash) are shown.  Each bin corresponds to $\sim0.38~\mu$secs.  The 
sampling interval for the top and bottom plots is $0.2\,\mu$secs, and for the
middle plot it is $0.1\,\mu$secs.  The 430~MHz data has been boxcar smoothed 
by 8 data points.  
\label{fig:BIG3}}
\end{figure}

The average giant pulse profiles constrain the properties of the giant
pulse emission region.  Also of interest are the properties of individual
giant pulses, which constrain the giant pulse emission mechanism itself.
To study individual giant pulses, which are narrower than the mean
giant pulse envelope, we must account for scattering effects 
even at frequencies above 430~MHz. As discussed above,
scattering by a thin turbulent screen degrades a signal by effectively 
convolving it with a one-sided exponential tail.  (Although more sophisticated 
models of the ISM will have more complex effects on the
observed signal, we find the simple thin-screen model adequately describes the 
vast majority of giant pulses we have observed.)  We have fit all candidate
giant pulses with a Gaussian,
$A\,\mbox{exp}\{-(t-t_0)^2/(2\sigma^2)\}$, convolved with an exponential
tail, $e^{-t/\tau}/\tau$, with four fit parameters, $A$, $t_0$, $\sigma$,
and $\tau$.  Shown in Fig.~\ref{fig:BIG3} are the strongest giant pulses
found at each frequency along with their fitted convolved Gaussians.

After visually inspecting many giant pulses at all three frequencies, we see no
evidence for intrinsic multiple-peaked emission, which contrasts with giant 
pulses observed from the Crab pulsar (\cite{sbh+99}).  Although receiver noise
dominates the profiles, most pulses show a fast rise followed by an
exponential decay, consistent, on average, with scattering.
We have more quantitatively verified our model for the pulse morphology by 
cross correlating all of the giant pulses with a standard giant pulse, 
shifting by the lag which maximizes the cross correlation (to correct for the 
observed pulse-to-pulse jitter), then folding.  
For any choice of the standard giant pulse, this has always produced a 
single-peaked, short-rise-time, exponential-decaying profile, which, itself, 
is well fit by our model.  Gaussian widths (FWHM) of $6$, $0.2$, 
and $0.2~\mu$secs and scattering timescales of $\tau\simeq29$, $0.2$, and 
$0.2~\mu$secs were determined at 430, 1420, and 2380~MHz, respectively, for
both the MP and IP shifted, folded giant pulse profiles.  As previously 
mentioned in \S\ref{sec:arr}, our exponential scattering model is probably an 
oversimplification of the true effects of scattering, implying that our 
estimates should be considered upper limits.

Table \ref{tab:cum} lists the range of scattering timescales, $\tau$ 
[$\mu$sec], found at each frequency.  Overall, we find the following ranges:  
$\tau\simeq13$--$40~\mu$s (430~MHz),  $\tau\lsim1.1~\mu$s (1420~MHz), and 
$\tau\lsim0.4~\mu$s (2380~MHz).  These timescales are consistent with 
turbulent scattering, which has a frequency dependence of 
$\nu^{-4\mathrm{\;to\;}-4.4}$ (e.g., \cite{mt77}).  In addition, we have 
determined approximate upper limits to the fitted Gaussian widths (FWHM) at 
each frequency of $\lesssim7~\mu$s (430~MHz), $\lesssim0.5~\mu$s 
(1420~MHz), and $\lesssim0.3~\mu$s (2380~MHz), which have not, however, 
been included in Table~\ref{tab:cum}.

Scattering also affects the normal emission, primarily observations at lower 
frequencies.  The best-fitting scattering parameters were determined 
independently in the two pulse components.  In the 
MP, our model finds $\tau\simeq30~\mu$sec, and in the IP,
$\tau\simeq40~\mu$sec.  Both are in 
good agreement with the scattering time estimated from the giant pulse 
emission.  The IP is well fit by this model, but our fit to the MP 
underestimates the flux in the tail, probably implying the unresolved presence 
at this frequency of the ``notch'' feature that is resolved on the trailing 
edge of the MP at higher frequencies.  (This is not unexpected, since the 
feature increases in strength relative to the MP peak with decreasing 
frequency.)  Note that scattering not only broadens but also delays the peak 
of the low-frequency
pulsar signal by an amount that depends on the pulse shape; variability of the 
scattering strength therefore introduces significant timing errors at low 
frequency.
For the 1420 and 2380~MHz normal emission, scattering is much less severe;
therefore, we expect no significant scattering delay to the apparent MP and IP 
peak arrival times.

\section{Timing}\label{sec:tim}

High-precision timing measurements benefit from a sharp-edged timing signal.
For this reason, the timing properties of the giant pulses are of considerable
importance, especially since very high signal-to-noise ratios can be
obtained for mean giant pulse profiles by using a signal thresholding
technique to eliminate data when no giant pulse is present.

\begin{figure}[c]
\centerline{\psfig{file=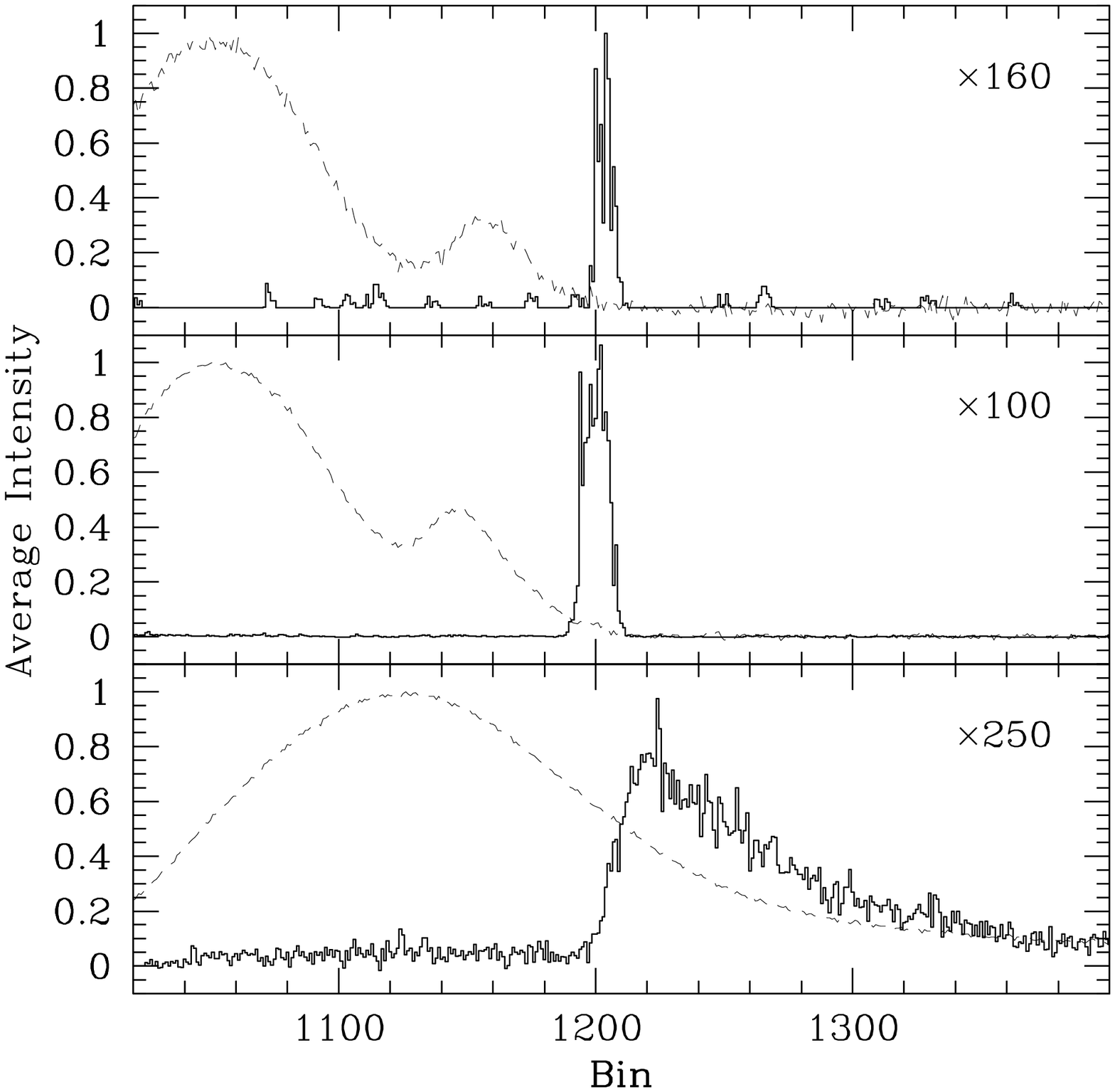,width=3.25in}}
\caption{Average giant pulse emission (solid) and normal emission (dashed) 
for the main-pulse region.  All other details are the same as for 
Fig.~\ref{fig:toothpick}.
\label{fig:toothpickMP}}
\end{figure}

\begin{figure}[c]
\centerline{\psfig{file=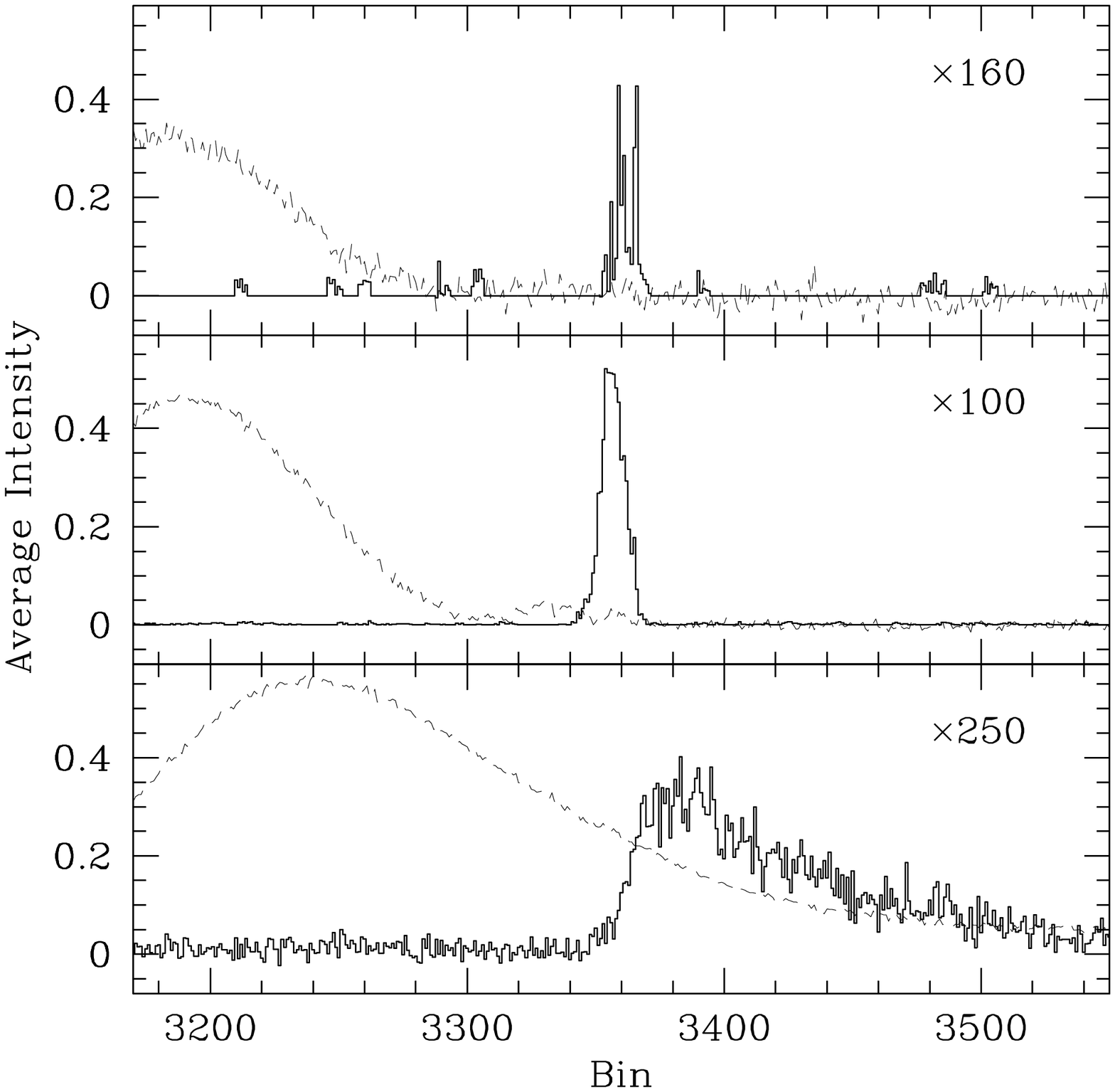,width=3.25in}}
\caption{Average giant pulse emission (solid) and normal emission (dashed) for 
the interpulse region.  All other details are the same as for 
Fig.~\ref{fig:toothpick}.
\label{fig:toothpickIP}}
\end{figure}

High-precision alignment of pulsar profiles at different frequencies is not 
straightforward when the pulse shape is variable, since the choice of fiducial 
reference phase for alignment is arbitrary.  We have aligned the three 
profiles in Figs.~\ref{fig:toothpick}--\ref{fig:toothpickIP} by the peaks of 
the normal emission profiles, in the case of 430~MHz after accounting for the 
delay introduced by scattering.  As is evident, at each frequency the giant 
pulses occur at approximately the same phase relative to the normal emission 
peaks.  The slight delay at 2380~MHz and possibly at 430~MHz with respect to 
the 1420~MHz giant pulse profile is most likely due to our somewhat 
\emph{ad hoc} alignment procedure.

We present timing characteristics for each observation in Table \ref{tab:cum}. 
Displayed for each scan are the separation in phase angle of the IP peak 
following the MP peak (Normal), the separation of the IP giant pulses from the 
MP giant pulses (Giant), and the delay (in [$\mu$s]) of the giant 
pulses with respect to the MP and IP emission peaks (with scattering taken 
into account for the 430~MHz observations, as will be discussed below).  We 
now discuss each timing column in more detail.

The separation of $57$--$58~\mu$secs between MP peak and average giant pulse  
(though only a lower limit of $49~\mu$secs at 430~MHz), as well as that for 
the IP of $65$--$66~\mu$secs, is the same at all three frequencies.  
This yields tight constraints on the relative geometry of normal and giant 
pulse emission regions.  The separation between the MP and IP giant pulses is 
$189.5^\circ$ at all frequencies, slightly larger than the $187.6^\circ$ 
separation the MP and IP normal emission peaks (though only a lower limit of 
$185.6^\circ$ can be quoted at 430~MHz).  The individual pulse arrival phases 
are Gaussian distributed, with widths of $\sigma=1.5$--$2.0~\mu$s, in good 
agreement with the width found for the average giant pulse profile (displayed 
in Figs.~\ref{fig:toothpickMP} and \ref{fig:toothpickIP}).  This 
pulse-to-pulse jitter is evident in Fig.~\ref{fig:arr}, where we have plotted 
the fractional giant pulse arrival bin versus pulse number for a 1420~MHz
observation.

It is interesting to ask whether observations of the giant pulse emission
from \psr\ can be used to carry out higher precision timing studies of
the pulsar than have been possible using the relatively broad normal
emission profile (e.g., \cite{ktr94}).  Using normal pulse timing
techniques, absolute precisions as small as $0.12~\mu$s have been
obtained for B1937+21 at 1420~MHz (\cite{sta98}).  Just as typical long term
timing studies depend on long-term stability of the normal emission profile,
timing studies using the giant pulses will depend upon long-term stability
of the giant pulse emission phase distribution.  The consistency of our
results with those of Cognard \etal\ are encouraging, but careful
observations at high frequency over a period of years will be needed
to test the ultimate power of giant pulse observations for high precision
timing.

\begin{figure}[c]
\centerline{\psfig{file=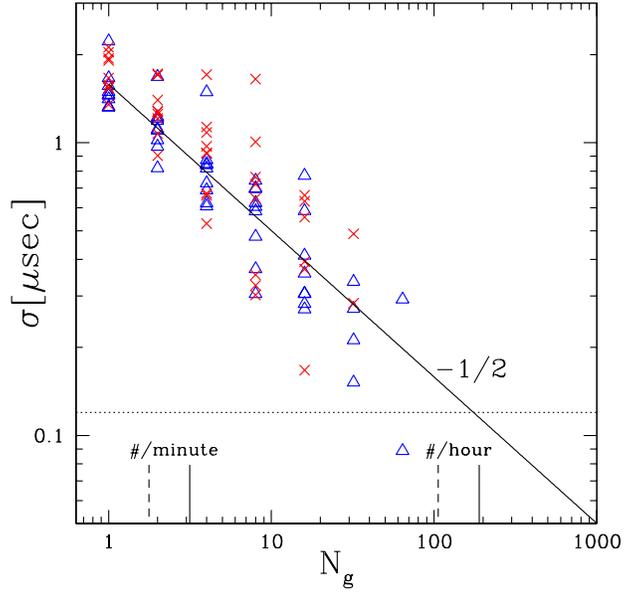,width=3.25in}}
\caption{Estimated giant pulse timing precision.  At each epoch where 1420~MHz 
observations were available, the standard deviation $\sigma$ of the giant 
pulse arrival phase is calculated for MP giants (triangles) and IP giants 
(X's), without averaging ($N_g=1$).  Then consecutive points are pairwise 
averaged ($N_g=2$) and $\sigma$ is calculated again.  Pairwise averaging 
continues as long as the number of points remaining is greater than one.  The 
solid line with slope $-1/2$ is expected for uncorrelated Gaussian timing 
noise, i.e., $\sigma\propto N_g^{-1/2}$.  We also plot the number of 
detectable giant pulses expected in 1-minute and 1-hour intervals for the MP 
(solid tick) and IP (dashed tick).  The dotted line is the timing precision, 
$\sigma\simeq120$~ns, achieved with the Mark~IV instrument using standard 
average-profile techniques (Stairs \etal\ (1999)).
\label{fig:sigma}}
\end{figure}

Nevertheless, we have done preliminary estimates of the obtainable timing
precision using the current data set.  As noted above, a single giant pulse at 
1420 MHz can be used to estimate the pulsar phase to $\sim1.5~\mu$s.  The key 
question for timing is whether this precision can be improved by averaging 
over multiple pulses.  If the pulse-to-pulse phase variations are uncorrelated,
we expect the timing precision to improve as $\sqrt{N_g}$, where
$N_g$ is the number of consecutive points averaged.  In Fig.~\ref{fig:sigma} ,
we show the r.m.s.\ scatter $\sigma$ within individual days as a function
of $N_g$.  We find that timing precision improves as expected to the
limit of our data sets, $N_g=64$, corresponding to a level of
100--300~ns in a ten-minute observation.  Also indicated in the figure
is the best timing precision achieved using standard pulsar timing
techniques, $\sim120$~ns.  Stairs \etal\ (1999) suggest that the
limiting factor in their timing analysis may be pulse shape variations
caused by variable interstellar scattering.  If this is correct,
then giant pulse observations may ultimately improve on normal
pulse observations, because the effects of scattering can be much
more easily measured and removed from the data.  Whether these
high-precision timing results on short timescales will lead to
better long term timing depends primarily on the long term stability
of the giant pulse emission characteristics, which will be studied
in future work.

\section{Intensity Distribution}\label{sec:int}

\begin{figure}[c]
\centerline{\psfig{file=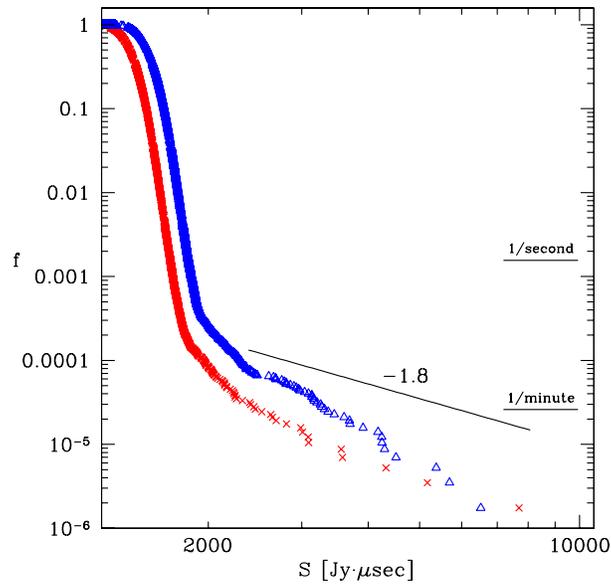,width=3.25in}}
\caption{Cumulative intensity distribution for all 150~$\mu$sec windows 
after the MP (triangles) and IP (X's) at 430~MHz (15 min.-run on MJD 51364), 
where $f$ denotes the fraction of pulses containing emission $\geq S$ in the 
specified windows.  Because the giant pulses arrive within the normal emission 
window, the giant pulses are augmented by the normal emission.  This steepens 
the observed giant pulse power law, leading to a poor fit to the displayed 
$-1.8$ power law.
\label{fig:cum430}}
\end{figure}

\begin{figure}[c]
\centerline{\psfig{file=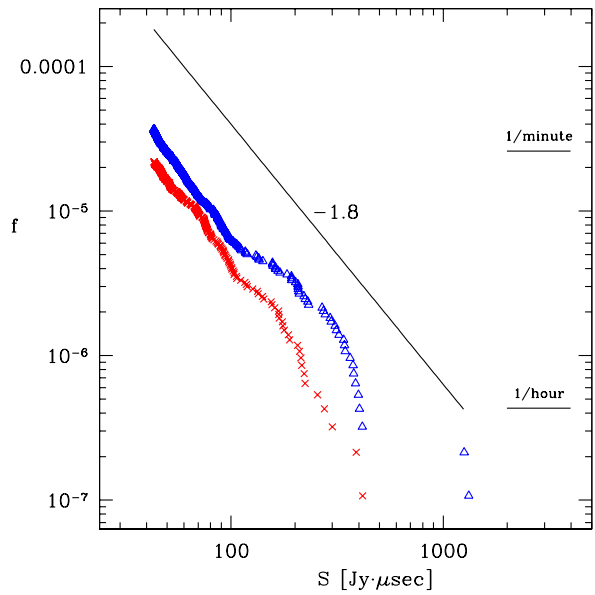,width=3.25in}}
\caption{Cumulative intensity distribution for candidate giant pulses at 
1420~MHz using the entire 4 hour data set, where $S$ is calculated over 
2~$\mu$secs for each candidate and $f$ denotes the fraction of pulses 
containing candidates arriving in the specified 11.4--$\mu$sec MP (triangles) 
or IP (X's) windows with intensity $\geq S$.  Also shown is a power law with 
index $-1.8$.
\label{fig:cum1420}}
\end{figure}

\begin{figure}[c]
\centerline{\psfig{file=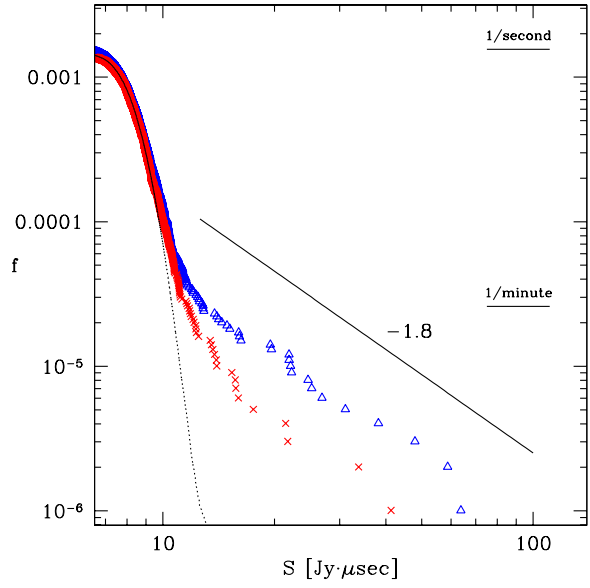,width=3.25in}}
\caption{Cumulative intensity distribution for candidate giant pulses at 
2380~MHz (26-min. run on MJD 51391), where $S$ is calculated over 2~$\mu$secs 
and $f$ denotes the fraction of pulses containing candidates arriving in the 
specified 15.2--$\mu$sec MP (triangles) or IP (X's) windows with intensity 
$\geq S$.  The dotted line denotes all other candidates from the rest of the 
pulsar phase scaled to the giant pulse window width of 15.2$~\mu$secs.  Also 
shown is a power law with index $-1.8$.
\label{fig:cum2380}}
\end{figure}

The primary distinguishing characteristic of the giant pulses observed
from the Crab Pulsar and \psr\ is, of course, their intensity,
and particularly 
their extended power-law intensity distribution.  In 
Figs.~\ref{fig:cum430}--\ref{fig:cum2380}, we plot the cumulative
intensity distribution at each observing frequency.  At low intensities, the 
distributions are dominated by the chi-square statistics of the noise and/or
normal emission, but above a certain threshold the pulse strength distribution 
is roughly power-law distributed.  Given our limited statistics for any given 
run, this simple model is adequate to describe the observed data.  

In Fig.~\ref{fig:cum430}, we plot the cumulative distribution of the 
integrated flux densities in $150~\mu$sec windows after the MP and IP during 
a single 15-minute observation at 430~MHz (MJD 51364).  Also plotted is the 
$-1.8$ power-law slope, which Cognard \etal\ (1996) found for the cumulative
giant pulse distribution at this same frequency.  The mean signal-to-noise 
ratio for the {\it normal} emission peak in this observation was about $0.14$. 
Contamination from this emission causes a noticeable deviation at low flux 
levels from what would be seen with giant pulses and receiver noise alone, 
causing a steepening of the distribution towards low integrated flux densities.
Although the distribution of normal pulse signal strengths is not known,
we expect that removing normal pulses would produce better accord with a 
single power law distribution for the giant pulses.  In 
Fig.~\ref{fig:cum1420}, we have plotted the cumulative distribution of all 
giant pulses found at 1420~MHz for all $\sim4~$hours of data (using the 
threshold detection algorithm described above).  Again, generally power-law 
behavior is observed, with a similar power-law exponent around $-1.8$. In 
Fig.~\ref{fig:cum2380}, we plot the cumulative intensity distribution for a 
$\sim26$-minute 2380~MHz run (MJD 51391).  Again we have plot a $-1.8$ slope 
for comparison, which appears to fit the MP giant pulses and the most 
energetic IP giant pulses well, though more data are needed to strengthen this 
result.  

We have also calculated the fraction, $R$, of the total pulsar emission at 
each frequency that emerges in the form of giant pulses.  We find the following
ranges at each frequency for this fraction:  $0.15\%\leq R\leq9\%$ (430~MHz), 
$0.13\%\leq R\leq4\%$ (1420~MHz), and $0.10\%\leq R\leq1\%$ (2380~MHz), 
where the lower limits were determined directly from 
Figs.~\ref{fig:toothpick}--\ref{fig:toothpickIP} and the upper 
limits were determined from Figs.~\ref{fig:cum430}--\ref{fig:cum2380} by 
assuming a cumulative distribution with power-law slope of $-1.8$ over all 
intensities for both the MP and IP giant pulses.  From the folded normal 
emission alone, however, we can rule out large values for $R$ in these 
calculated ranges, implying that the single power-law model may 
not be valid at low intensities.

Also consistent with Cognard \etal\ (1996), we find significantly stronger 
giant pulses following the MP than the IP.  At a given frequency
(e.g., 1/minute), the ratio of the strongest giant pulse associated with the
MP to the strongest associated with the IP is very roughly the same as the 
ratio of the peak flux density in the MP to that in the IP.  Despite the fact 
that giant pulses are separated in pulse phase from the normal emission, this 
suggests a relatively close association between the emission processes.

\section{Spectrum}\label{sec:spec}

The short timescale of the observed giant pulses imply that they
are a relatively broadband phenomenon, with bandwidth greater than
their inverse width.  The limited bandpass available to the Mark~IV
instrument prevents stronger statements about the spectra of
individual giant pulses. However, the similar arrival distributions and roughly
similar arrival rates at each observational frequency point to a broadband 
phenomenon.

Nevertheless, our observations can be used to constrain the average spectral 
properties of the giant pulse emission.  To avoid complications arising from 
the use of different effective thresholds at each frequency (because of 
different source strengths and different receiver noise properties), we 
estimate the giant pulse spectrum by using the most powerful individual pulses 
observed during a given time period at each frequency.  We have, as usual,
calibrated our observations to the spectrum for the average normal emission 
flux density at each frequency from Foster \etal\ (1991), as discussed in 
\S\ref{sec:obs}.

\begin{figure}[c]
\centerline{\psfig{file=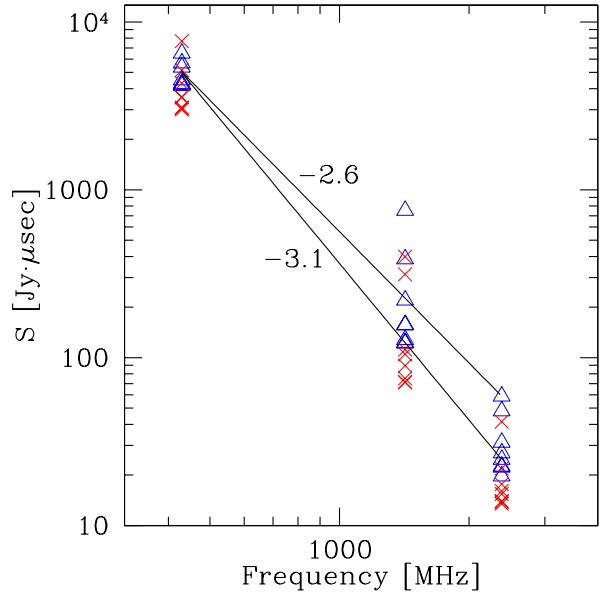,width=3.25in}}
\caption{Giant pulse integrated flux density versus frequency for the eight 
strongest giant pulses in the MP and in the IP taken from equal-length scans 
of 15 minutes on MJD 51364 (430~MHz), 50893 (1420~MHz), and 51391 (2380~MHz).  
At each frequency, the absolute flux density calibration was calculated 
independent of scintillation using the spectral model of Foster \etal\ 
(1991).  This power-law model has a spectral index for normal emission 
of $-2.6$; that slope is indicated above.  There is some evidence that the 
spectrum of the strongest giant pulses is steeper; the best fit power-law 
model, with index $-3.1$, is also shown.
\label{fig:spec}}
\end{figure}

Figure \ref{fig:spec} shows the intensities in [Jy$\cdot\mu$sec] of the top 
eight MP and top eight IP giant pulses at each 
frequency over 15~minutes, corresponding to the entire MJD 51364 (430~MHz) 
run and to 15-minute chunks from runs on MJD 50893 (1420~MHz) and MJD 51391 
(2380~MHz)  We find a somewhat steeper slope of $-3.1$ for the 
giant pulse spectrum, compared to the $-2.6$ slope for the normal emission
spectrum.  Although the precise slope of the giant pulse emission spectrum
depends on the assumed normal emission spectrum, the result that 
the giant pulse emission is steeper than the normal emission is robust.  
However, if the apparent narrowing of the giant pulse emission region at 
higher frequency reflects a narrowing of a sharp emission cone, the slightly 
steeper spectrum of the giant pulse emission might be understood as the 
geometric effect of the position of the line of sight through the outer part 
of the emission region

Assuming the giant pulses from the Crab pulsar are powered by
curvature radiation, Sallmen \etal\ (1999) have calculated the necessary
number and number density of radiating electrons.  We perform
the same calculation here for our observed giant pulses from \psr.
For coherent curvature radiation, the power emitted by $N$ electrons
with relativistic factor $\gamma=(1-v^2/c^2)^{-1/2}$ travelling
along magnetic field lines with radius of curvature $\rho_c$ is
\begin{equation} P_{curv}=N^2\bigg(\frac{2e^2\gamma^4c}{3\rho_c^2}\bigg).
\end{equation} 
From the peak of the largest 1420~MHz giant pulse
in Fig.~\ref{fig:BIG3}, we can calculate the maximum number of
electrons needed in one bunch to produce the observed emission (at 
frequencies greater than 430~MHz).
Assuming this pulse is broadband with spectral index, $-3.1$, \psr\
is at a distance of 3.6~kpc, and giant pulse beaming is determined
by the beam width $\theta\simeq\gamma^{-1}$, we find a total power
at the peak of $\gamma^{-2}\times10^{34}~$erg$\cdot$s$^{-1}$.  An upper
limit on the giant pulse Gaussian FWHM of $\lsim0.5~\mu$sec (at high
frequencies), gives $\gamma\gsim500$, which implies a substantially
smaller power requirement than the total spin-down energy loss
rate of $2\times10^{36}~$erg$\cdot$s$^{-1}$.  Setting $P_{curv}$ equal
to the observed power yields \begin{equation} N = 10^{19}
\bigg(\frac{\gamma}{500}\bigg)^{-3}
               \bigg(\frac{\rho_c}{10^6~\mathrm{cm}}\bigg).
\end{equation} In order to preserve coherence at the
highest observational frequency, these electrons must
fit within a cube with volume $\lsim\lambda^3$, where
$\lambda=12.6$~cm at 2380~MHz, implying an electon density of
$n_e=5\times10^{15}(\frac{\gamma}{500})^{-3}(\frac{\rho_c}{10^6~
\mathrm{cm}})~$cm$^{-3}$,  a value $50$ times greater than the 
Goldreich-Julian density (\cite{gj69}), 
$n_{{\mathrm{G-J}}}=\Omega B/2\pi e\,c=10^{14}\, (\frac{R}{R_{{\mathrm{NS}}}})^{-3}~$cm$^{-3}$, which is the electron number density required to power the
normal emission via curvature radiation.  Sallmen \etal\ (1999) 
similarly find a giant pulse electron density $100$ times greater (for 
$\rho_c=10^7$~cm) than the Goldreich-Julian value for the Crab 
pulsar.

\section{Discussion}\label{sec:dis}

The discovery of giant pulses from a millisecond pulsar was unexpected,
and early hopes that identifying commonalities between \psr\ and
the Crab pulsar might lead to a better understanding of the giant
pulse emission mechanism have so far not been realized.  As our study
has confirmed, the giant pulse emission from \psr\ differs
fundamentally from the Crab giant pulses, despite their common power-law
behavior.  The most intriguing characteristic of the high-frequency
pulses form \psr\ is their very narrow widths and the very limited regions
of pulse phase in which they occur.   

Despite the continued mystery about their origin, it appears likely that
giant pulses from \psr\ may prove a valuable tool.  As we have discussed,
their narrow intrinsic width and large flux make them attractive fiducial
reference points for timing studies of a pulsar that is already
among the most precisely timed (Kaspi \etal\ 1994).
Another intriguing possibility is to use the giant pulses as bright
flashbulbs to study scattering in the ISM.  Because the intrinsic pulses are
very narrow, the pulse shape as observed at the Earth traces out the
time delays introduced by multipath scattering.  The combination of this
information with VLBI studies of the scattering disk is a potentially
powerful tool for studying the three dimensional distribution of scattering
material.

It is important, of course, to identify giant pulse emission from other
pulsars.  As we have noted, the integrated giant pulse emission from \psr\
is too weak to make noticeable features in the average pulse profiles, so
careful, single pulse studies are required.  Very fine time resolution is
needed to avoid substantially smearing the high-frequency pulses from
B1937+21 and reducing their signal-to-noise ratio, and it is insufficient
to study only the windows of pulse phase where normal emission is found,
as has sometimes been the case in past studies with coherent dedispersion
instruments.  Instruments that use hardware dedispersion followed by
sampling (like the Princeton Mark III, \cite{skn+92}) must also preserve
sufficient dynamic range in the analog-to-digital conversion to detect
and characterize pulses that are far stronger than the typical pulsar
emission.  Another subtlety concerning observation of giant pulses are possible
projection effects.  If the \psr\ giant pulse emission region is roughly 
Gaussian, then as the pulsar rotates, the giant pulse emission traces out
less than $1\%$ of the entire sky, though the likelihood of detecting giant 
pulses from a known radio pulsar might be substantially enhanced over this
estimate by correlations between the angular patterns of the normal
and giant emission.  Although searches for giant pulse 
emission from slow pulsars have been unsuccessful, only a very small fraction 
of millisecond pulsars have been studied sufficiently to detect or rule out 
giant pulses.

\acknowledgments{A.K. would like to thank I. Stairs, in particular, for 
her assistance with the analysis, E. Splaver for his thorough explanations,
J. Taylor and J. Bell-Burnell for their observing efforts, and the other 
members of the Princeton Pulsar Group, especially D. Nice for constructive
comments on the draft.  In addition, S.E.T. thanks his earlier 
collaborators in this work, I. Cognard and J. Shrauner. This 
research was funded in part by a grant from the National Science Foundation,
which also supports A.K. through a graduate fellowship. Arecibo Observatory
is operated by Cornell University for the NSF.}

\clearpage
\bibliographystyle{apj}


\clearpage

\clearpage

\begin{deluxetable}{rrccccccc}
\scriptsize
\tablewidth{0pc}
\tablehead{\colhead{Freq.}&\colhead{Min.}&\colhead{Normal}&\colhead{Giant}&\colhead{MP delay [$\mu$sec]}&\colhead{IP delay [$\mu$sec]}&\colhead{$\tau$ [$\mu$sec]}}
\startdata
2380 &  25.8 & $187.9\pm0.3^\circ$ & $189.5\pm0.5^\circ$ & $58\pm1$ & $65\pm2$ & $\lsim0.4$  \\
1420 & 242.6 & $187.58\pm0.01^\circ$ & $189.5\pm0.2^\circ$ & $57\pm1$ & $65\pm1$ & $\lsim1.1$\\
430  &  30.0 & $\gsim 185.6\pm0.5^\circ$ & $189.4\pm0.6^\circ$ & $\gsim 49\pm2$ & $66\pm2$ & $13$--$43$\\
\enddata
\label{tab:cum}
\end{deluxetable}

\end{document}